\RequirePackage{fix-cm}
\documentclass[smallextended]{svjour3}       
\smartqed  
\usepackage{graphicx}
\usepackage{amsmath}
\usepackage{amssymb}
\usepackage{subfigure}
\usepackage{booktabs}
\usepackage{cite}
\usepackage{pifont}
\usepackage{multirow}
\usepackage{longtable}

\begin{document}

\title{A Scale-Free Topology Construction Model for Wireless Sensor Networks
}


\author{
	Lurong  Jiang      	\and
	Xinyu Jin        		\and
	Yongxiang Xia		\and
	Bo Ouyang        		\and
	Duanpo Wu		\and
	Xi Chen
}


\institute{
L. Jiang \and X. Jin(\ding{41}) \and Y. Xia \ and B. Ouyang \and D. Wu \at
              Department of Information Science \&  Electronic Engineering, ZheJiang University, HangZhou, China. \\
              \email{jinxy@zju.edu.cn}           
           \and
L. Jiang 	\at
              \email{jianglurong@zju.edu.cn}           
           \and
Y. Xia 	\at
              \email{xiayx@zju.edu.cn}           
           \and
B. Ouyang \at
              \email{ouyangbo@zju.edu.cn}           
           \and
D. Wu 	\at
              \email{wuduanpo@126.com}           
           \and
X. Chen \at
              State Grid Information \&  Telecommunication Co., Ltd., China   		\\
            \email{xichen@sgcc.com.cn}           
}

\date{Received: date / Accepted: date}

\maketitle

\begin{abstract}
A local-area and energy-efficient (LAEE) evolution model for wireless sensor networks is proposed.
The process of topology evolution is divided into two phases. In the first phase, nodes are distributed randomly in a fixed region.
In the second phase, according to the spatial structure of wireless sensor networks, topology evolution starts from the sink, grows with an energy-efficient preferential attachment rule in the new node's local-area, and stops until all nodes are connected into network.
Both analysis and simulation results show that the degree distribution of LAEE follows the power law. This topology construction model has better tolerance against energy depletion or random failure than other non-scale-free WSN topologies.
\end{abstract}

\section{Introduction}

Wireless Sensor Networks (WSNs) are a kind of self-organized distributed wireless networks composed of a large quantity of energy-limited nodes. Topology construction is one of the primary challenges in WSNs for ensuring network connectivity and coverage, increasing the efficiency of media access control protocols and routing protocols, improving the routing efficiency,  extending the network lifetimes, and enhancing the robustness of the network \cite{abbasi2013overview,Uster2011,Akbari2013,Xu2011,Li2006}. The main aim of topology construction is to build a topology to connect network nodes based on a desired topological property. A dense network topology leads to high energy consumption due to overlapped sensing areas and maintenance costs of topology, while a very sparse network topology is vulnerable to network connectivity \cite{Qureshi2013}.


The development of complex networks provides new ideas for topology construction of WSNs. Study on complex networks is a newly emerging subject that focuses on the networks which have non-trivial topological features \cite{Lu2009,Cui2010}. There are many common characteristics between WSNs and typical complex networks models: networks contain a large number of nodes and have non-trivial topological features, and nodes in networks connect to each other through multi-hop paths \cite{Wang2007}. More importantly, typical complex network models, such as the small-world \cite{Watts1998,Newman1999} and scale-free \cite{Barabasi1999} network models, show some characteristics which are beneficial in WSNs. Small-world networks present small average path length between pairs of nodes which is beneficial to saving energy in topology construction and routing in WSNs \cite{Guidoni2010}. Scale-free networks have power-law degree distributions, and show an excellent robustness against random node damage \cite{Albert2000,Gao2010}. A random attack does not significantly affect the scale-free network performance \cite{Cui2010,Xia2008}. Therefore, it is significant to consider complex networks topology when optimising the topology in WSNs \cite{MIshizuka2004}. However, complex networks are a kind of relational graphs whose nodes make direct contact according to their logical relationships, while WSNs are spatial graphs in which the existence of links depends on the node's positions and radio range \cite{Helmy2003}. Thus, the complex networks theory cannot be directly used in WSNs. Some efforts have been made to tune wireless networks into heterogeneous networks with small-world \cite{Guidoni2010,Chitradurga2004,Cavalcanti2004,Verma2011,Ye2008} or scale-free features \cite{luo2011energy,Wang2007,Xuyuan2009,Hailin2009}.

In this paper, we propose a local-area and energy-efficient (LAEE) evolution model to build a WSN with scale-free topology. In this model, topology construction is divided into two phases. In the first phase, nodes are distributed randomly in a fixed region, and a node gets other node's information in its radio range through HELLO message. In the second phase, topology evolution starts from the sink, grows with preferential attachment rule, and stops until all nodes are added into network. Following conditions are considered when we design the evolution model: (i) Links between nodes depend on the positions and transmission range (radio range). Therefore, nodes beyond transmission range cannot make direct contact. (ii) Nodes can only get local information as WSNs are distributed networks. (iii) The remaining energy of each node is considered. Nodes with more remaining energy have higher probability to be connected. (iv) In order to avoid excessive energy consumption, upper bound of degree for each node is needed.

The remainder of this paper is organized as follows: Section 2 reviews background and related works on scale-free networks and scale-free based wireless networks. In section 3, we propose the LAEE evolution model, and deduce the theoretical degree distribution. Section 4 shows simulation results based on LAEE evolution model, and examines the tolerance of LAEE to random failures. Finally, we conclude in section 5.

\section{Background and Related Work}

\subsection{Traditional Topology Constructions in WSNs}

Unit disk graph (UDG) is the underlying topology model for WSNs which contains all links in transmission range (radio range).  Assume that all nodes are randomly distributed in region $ S $. Each node is positioned in a particular subarea with independent probability $ \varphi = \pi r^2 / S $, where $ r $ is the transmission range. The probability that a subarea has $ k $ nodes is given by the binomial distribution, $ p(k) = ( \begin{array}{r} n \\ k \end{array} ) \varphi ^k (1- \varphi)^{n-k} $, where $ n $ is the total number of nodes in the network. With the increase of $ n $, this probability becomes the Poisson distribution $ p(k) = (n \varphi)^k e^{-n\varphi}/k!  $. Then the average number of neighbor nodes is close to $ n \varphi -1 $. However, UDG model has high concentration of connections that might promote excess energy consumption for periodic topology maintenance and route selection process. Therefore this is an inefficient way of topology construction.

Almost all other topology construction methods in WSNs build a reduced topology from UDG \cite{Wightman2011,SJardosh2008}. Based on the topology production mechanism, they can be categorized into \textit{Flat Networks} and \textit{Hierarchical Networks} with clustering\cite{SJardosh2008}.

In \textit{Flat Networks}, all nodes are considered to perform the same role in topology and functionality. Typical examples include directed relative neighborhood graph (DRNG) \cite{NLi2004}, k-nearest neighbor (KNN) \cite{DMBlough2006}, TopDisc\cite{BDeb2002}, Euclidean minimum spanning tree (EMST) \cite{PKAgarwal1991}, local Euclidean minimum spanning tree (LEMST) \cite{NLi2003}, Delaunay triangulation graph (DTG) \cite{MLi2003}, and cone-based topology control algorithm(CBTC) \cite{LLi2005}.

In KNN, a node sorts all other nodes in its transmission range in Euclidean distance or other distance metric, and then links the $ k $ nearest nodes as neighbors in the final topology. It is a scalable and parameter-free in WSNs and very easy to implement. In DRNG, a link connects node $ u $ and $ v $ if and only if there does not exist a third node $ w $ that closer to both $ u $ and $ v $ in distance. TopDisc discovers topology by sending query messages and describing the node states using three or four color system. It is a greedy approximation method based on minimum dominating set. In EMST or LEMST, each node builds its overall or local minimum spanning tree based on Euclidean distance and only keeps nodes on tree that is one hop away as its neighbors. In DTG, a triangle formed by three nodes $ u $, $ v $, $ w $ belongs to topology if there is no other node in the scope of the triangle. CBTC uses an angle  $ \alpha $ as a key parameter. In every cone of angle $ \alpha $ around node $ u $, there is some node that $ u $ can reach.

Nodes in \textit{Hierarchical Networks} with clustering are heterogeneous in functionality as cluster heads or cluster members. LEACH is a typical \textit{Hierarchical Network} topology model \cite{WBHeinzelman2002} in which the network is clustered and periodic updated. The cluster heads have the responsibility to communicate directly with the sink for the whole cluster members. A node selects itself to be a cluster head with a probability related to factors such as its remaining energy, and whether it has served as cluster head in the last $ r $ rounds.

The WSNs topology can be indicated as graph $ G(V, E) $, where the sets of $ V $ and $ E $ are sensor nodes and topological links, respectively. We denote the number of a node's links, also the number of its neighbor nodes, as its degree. All these previous topology construction models show highly concentrated degree distribution, which means these models tend to present homogeneous graph property \cite{CMa2011,CTong2012}.

\subsection{Scale-free Evolution Models}
Barab\'{a}si and Albert provide an evolution model, called BA model, to generate a scale-free network. This model includes the following two features. (i) Growth: The network starts with a small number $ m_0 $ of nodes. At each time step a new node with $ m  (m \leqslant  m_0) $ edges is added. (ii) Preferential attachment: The new node connects to existing node $ i $ according to the probability $ \Pi_i = k_i / \sum_j k_j $, where $ k_i $ is the degree (i.e., numbers of topological links) of node $ i $. In BA model, the degree distribution follows the power-law distribution $ P(k) = k^{- \gamma} $, where the scaling exponent is $ \gamma = 3 $. The BA model cannot be directly used to generate a WSN because the overall network's degree $ \sum_j k_j $ is needed in BA model, which is unable to achieve in many real networks. As the limitations of transmission range, energy, and processing capacity, nodes in WSNs can only collect information from n-hop neighbor nodes but cannot get global information.

Li and Chen propose a local-world evolution model \cite{XiangLi2003}. In local-world evolution model, the preferential attachment does not work on the global network, but works on a local world of each node. $ M $ nodes are randomly selected from existing network as the \textit{local world} for the new node. The preferential attachment probability for new node at time step $ t $ is

\begin{eqnarray}
\label{formula1}
 \Pi_i = \Pi'(i  \in local-world) \frac{k_i}{\sum\limits_{j \in local-world} k_j}
\end{eqnarray}

\noindent where $ \Pi ^{'} (i  \in local-world) = M / (m_0 + t) $. As these $ M $ nodes are selected randomly, the spatial relationships between nodes are not considered. Therefore, the local-world evolution model still cannot describe the topology evolving mechanism in wireless networks.

\subsection{WSNs Topology Constructions with Scale-free Theory}
Several methods have been proposed to build WSNs with the scale-free property \cite{CTong2008,Xuyuan2009,Wang2007,Hailin2009}. These methods take complex network characteristic such as growth, preferential attachment into account, and some of them consider the local-area feature in WSNs.

Zhang provides a model of WSNs based on scale-free network theory \cite{Xuyuan2009}. In this model, each node has a saturation value of degree, $ k_{max} $, to balance energy consumption. The newly generated node has a certain probability $ P_e $ to be damaged when it is being added to the network. The probability that the new node will be connected to the existing nodes as follows:

\begin{eqnarray}
\label{formula2}
\Pi_i = P( d_{iv} \leqslant r)(1 - P_e) \frac{k_i}{\sum\limits_{total-network}{k_j - q k_{max}}}
\end{eqnarray}

\noindent where $ d_{iv} $ is the distance between the new node $ i $ and existing node $ v $, $ r $ is the transmission range, and $ q $ is the number of nodes which already reach the saturation value of degree $ k_{max} $. In Eq. \eqref{formula2},  $ P(d_{iv} < r) $ refers to the ratio of $ \pi r^2 $  to $ S $, where $ S $ is the entire WSNs coverage region.

One of the main problems of Zhang's model is the sum of $ \Pi_i $  is much smaller than 1. The scaling exponent is $ \gamma = 1 + 2S / \pi r^2 $, where $ S $ is the entire coverage region and $ r $ is the transmission range. Therefore, another problem is that the scaling exponent $ \gamma $ of degree distribution is much greater than 3, which is not rational in real networks.

Wang et al. propose an arbitrary weight based scale-free topology control algorithm (AWSF) \cite{Wang2007}. All nodes in the network are coupled with a sequence of random real numbers $ W $ with a power-law distribution $ \rho (x) = A x^{-\theta} $, where $ \theta > 1 $ and $ A = \int_{min}^{max} \rho (x) d x = 1 $. The balance of energy consumption is not considered in this model. Therefore, there is a possibility that a node with low energy coupled with a large weight $ w $ and therefore has a large degree, which exacerbates the imbalance of energy consumption.

Zhu proposes an energy-aware evolution model (EAEM) of WSNs \cite{Hailin2009}. Energy is taken into account in the EAEM model. This algorithm assumes that the probability $ \Pi_i $ that a new node connects to existing node $ i $ depends on its degree $ k_i $ and the remaining energy of that node. A function $ f(E) $ is defined to present the relationship between remaining energy and its ability to be linked. $ f(E) $ must be an increasing function, as the more energy a node has, the more probability it will be connected to the new node. Therefore the form of $ \Pi_i $ is

\begin{eqnarray}
\label{formula3}
\Pi_i = \frac{f(E_i)k_i}{\sum\limits_{local-area} f(E_j)k_j}
\end{eqnarray}

\noindent where the \textit{local-area} in the EAEM is the set of nodes locating in the new node's transmission range. The sum of all nodes'  $ \Pi_i $ is less than 1 and the scaling exponent $ \gamma $ of degree distribution is 1, which is not rational in real networks.

\section{Local-area and Energy-efficient Evolution Model}

In this section, we propose our scale-free topology construction model for WSNs.

\begin{figure}
  \includegraphics[width=0.6\textwidth]{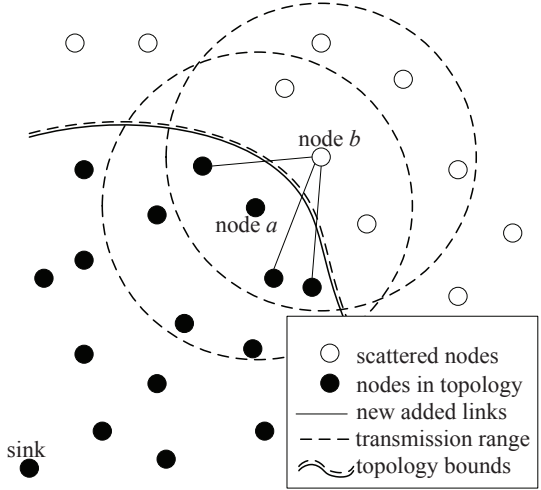}
  \caption{At every time step, add a scattered node into the network with $ m (m<m_0) $ links}\label{Fig1}
\end{figure}

Usually, nodes are distributed in a given region with static positions. Then connections between them are built to generate a network. Based on this fact, the process of topology construction is divided into two phases: In the first phase, nodes are distributed randomly. We define the set of \textit{scattered nodes}  as the nodes having not access to the network topology yet in the process of evolution, as shown in Fig. \ref{Fig1}. An arbitrary node, marked as node $ v $, gets all other nodes' information in its transmission range through HELLO message, and takes these nodes as its  \textit{potential neighbor nodes}. Then in the second phase, topology evolution starts from sink, grows with the preferential attachment rule, and stops until all nodes are added into the network.

The LAEE evolution model is proposed:

\begin{quote}

Step I.
\begin{quote}
Nodes are distributed randomly in region $ S $. Each node gets its potential neighbor nodes' information in its transmission range through HELLO message. All these nodes are scattered and topology has not been formed at this moment.
\end{quote}
Step II.
\begin{quote}
II.1\ \ Topology evolution starts from sink with $ m_0 $ nodes (the sink and its $ m_0 - 1 $ potential neighbor nodes) and $e_0$ random links between them.

II.2\ \ At every time step, add a scattered node into the network. To do that, we find the node which has the most scattered neighbor nodes, and mark it as node $ a $. Choose a scattered node randomly in node $a$'s potential neighbors as the new node, denoting as node $ b $. With this strategy, the network expands outward and fills the region $ S $ as fast as possible.

II.3\ \ Randomly choose $m$ nodes, which are already in the topology and are node $b$'s potential neighbors, and link them to node $b$. If the number of node $b$'s potential neighbors is smaller than $ m $, all these nodes will be linked to this new node. Connect node $ b $ with $ m $ potential neighbor nodes based on the preferential attachment:

\begin{eqnarray}
\label{formulaP}
\Pi_i = \Pi'_i (i \in {local-area}) \frac{f(E_i)k_i}{\sum\limits_{local-area}{f(E_j)k_j-q k_{max}}}
\end{eqnarray}

\noindent where \textit{local-area} is the set of node $b$'s potential neighbor nodes in its transmission range, $ k_{max} $ is the upper bound of node's degree, $ q $ is the number of nodes which already have the degree of $ k_{max} $, and $ f(E) $ is the function mentioned in the EAEM model. When a node reaches the degree of $ k_{max} $, no more link can be added to it. 

II.4\ \ Repeat II.1,II.2, and II.3 until all nodes are added to the topology.
\end{quote}
\end{quote}

In Eq. \eqref{formulaP}, $ \Pi'_i(i \in {local-area}) $  refers to the set of node \textit{i}'s neighbor nodes in its transmission range at time step $ t $, i.e.,

\begin{eqnarray}
\label{formula5}
\Pi'_i (i \in {local-area}) = {n \varphi}/{(m_0 + t)}
\end{eqnarray}

\noindent where $ n $ is the number of nodes in network, $ \varphi = \pi r^2 / S $ is the possibility of two nodes positioned in each other's transmission range. We assume that only few nodes reach the upper bound $ k_{max} $, so $ qk_{max} $ could be ignored here. Therefore, in local area we have

\begin{eqnarray}
\label{formula6}
\sum\limits_{local-area}{f(E_j)k_j - q k_{max}}  & \approx & \sum\limits_{local_area}{f(E_j)k_j}
 =  {\overline M \  \overline E <k>}
\end{eqnarray}

\noindent where $ \overline M $ is the expected number of nodes in new node's local-area which equal to $ n\varphi $ as the expected number of nodes in transmission range mentioned in UDG model, $ \overline E $ is the expected value of $ f(E) $, and $<k> = 2(mt + e_0)/(m_0 + t) $ is the average degree of network at time step $ t $, where $ m_0 $ and $ e_0 $ represent the number of nodes and links at the beginning, respectively. Then we get the varying rate of $ k_i $:

\begin{eqnarray}
\label{formula7}
\frac{\partial k_i}{\partial t} & = & m \Pi_i 										\nonumber\\
& = & m \Pi'_i(i \in {local-area}) \frac{f(E_i)k_i}{\sum\limits_{local-area}{f(E_j)k_j - q k_{max}}}	\nonumber\\
& = & m \frac{n \varphi}{m_0 + t}\frac{f(E_i)k_i}{n \varphi \overline E {\frac{2(mt+e_0)}{m_0+t}}}
=\frac{m f(E_i) k_i}{2 \overline E (mt + e_0)}
\end{eqnarray}

In a very large scale network, $ e_0 $ can be ignored, then we can get

\begin{eqnarray}
\label{formula8}
\frac{\partial k_i}{\partial t}
\approx
\frac{f(E_i)k_i}{2 \overline E t}
\end{eqnarray}

As $ f(E) $ is an increasing function, we set $ f(E_i)k=E $, Then

\begin{eqnarray}
\label{formula9}
\frac{\partial k_i}{\partial t} = \frac{E k_i}{2 \overline E t}
\end{eqnarray}

According to the initial degree of node $ i $ at time step $ t_i $, $ k_i(t_i)=m $, we can get

\begin{eqnarray}
\label{formula10}
k_i = m {(\frac{t}{t_i})} ^ \beta
\end{eqnarray}

\noindent where $ \beta = E / {2 \overline E} $. The probability that node \textit{i}'s degree is smaller than $ k $ is

\begin{eqnarray}
\label{formula11}
p(k_i(t) <  k) & = & 1 - p(t_i < t(\frac{m}{k}) ^ {1/ \beta}) = 1 - \frac{t(\frac{m}{k})^{1/ \beta}}{m_0 + t}
\end{eqnarray}

Then we can obtain the probability density of the degree of a node with energy $ E $ as

\begin{eqnarray}
\label{formula12}
P(k_E)
& = & \frac{\partial p(k_i(t) < k)}{\partial k}
= {\frac{1}{\beta}} m^{1 / \beta} {\frac{t}{m_0 + t}}k^{-(1+1 / \beta)}				\nonumber\\
& \approx &  {\frac{1}{\beta}} m^{1 / \beta} k^{-(1+1 / \beta)}
\end{eqnarray}

In the above equation, $ \beta \in (E_{min}/2 \overline E, E_{max}/2 \overline E) $, where $ E_{min} $ and $ E_{max} $ are the bounds of energy $ E $. Therefore, the distribution $ P(k_E) $ has a power-law from with degree exponent  $ \gamma = (1 + 1/ \beta) $.

In order to get the probability density of degree with remaining energy $ E $, we have
\begin{eqnarray}
\label{formula13}
P(k)
& = & \int_{E_{min}}^{E_{max}} \rho P(k_E) d E 	 = \int_{E_{min}}^{E_{max}} \rho {\frac{1}{\beta}} m^{1/ \beta}k^{-(1+1/ \beta)} d E
\end{eqnarray}

\noindent where $ \rho $ is the distribution of $ E $ with the bounds of $ E_{min} $ and $ E_{max} $.  $ \rho $ satisfies the equation
$ \int_{E_{min}}^{E_{max}} \rho d E = 1 $.

\section{Numerical Results}

Table \ref{table1} presents the parameters used in our simulation. We distribute $ n=1000 $ nodes randomly in the square region $ S $, and deploy the sink at a corner marked as (0, 0). We select $m_0=10$ nodes and $e_0=10$ links in sink's transmission range as the initial state of our evolution model. Energy in the networks are uniformly distributed. The value of $ \rho $ is a constant which can get from the equation $ \int_{E_{min}}^{E_{max}}\rho dE = 1$. Different values of $ m $ in LAEE are considered in our simulation.

\begin{figure}[!htbp]
\subfigure[$ m = 3 $]{
\label{Fig2a} 
\includegraphics[width=0.48\linewidth]{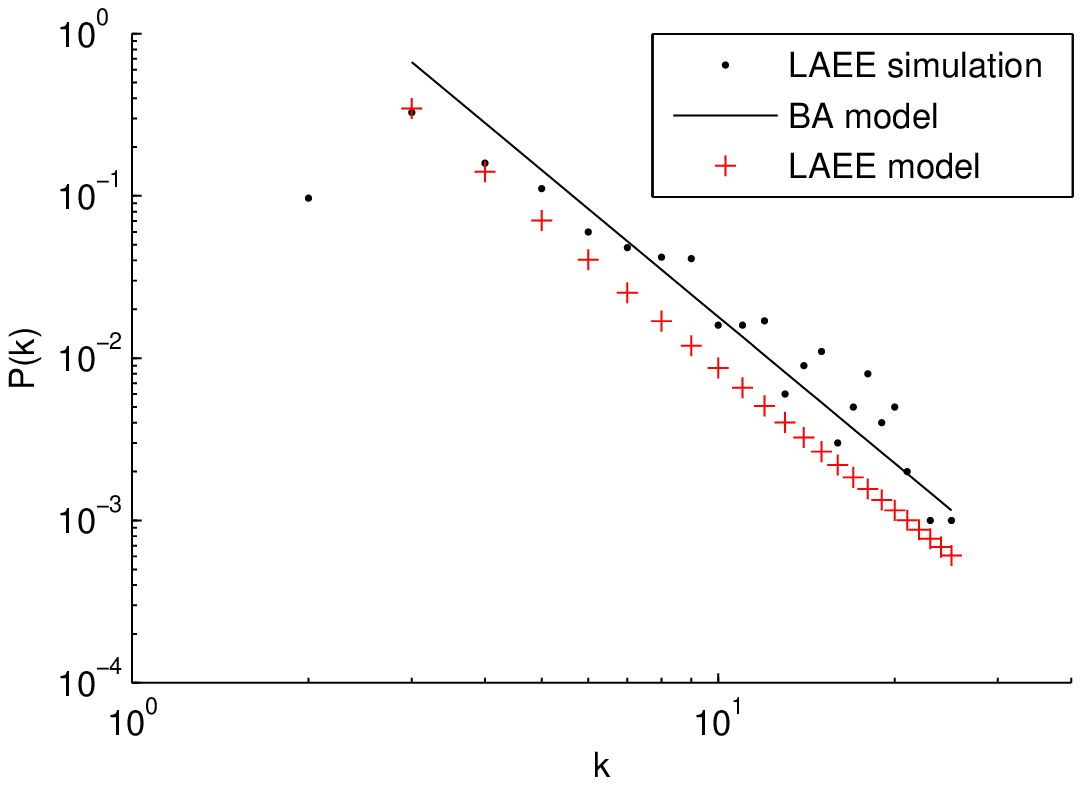}}
\hspace{0.01in}
\subfigure[$ m = 5 $]{
\label{Fig2b} 
\includegraphics[width=0.48\linewidth]{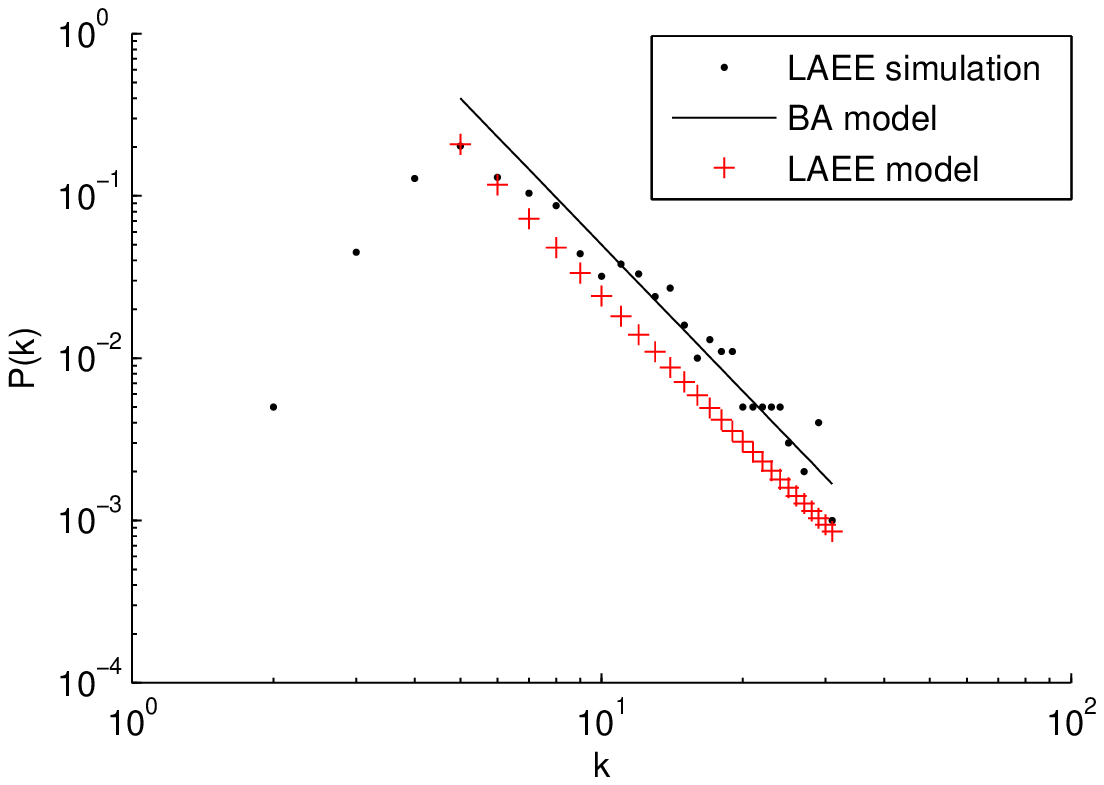}}
\hspace{0.01in}\\
\subfigure[$ m = 8 $]{
\label{Fig2c} 
\includegraphics[width=0.48\linewidth]{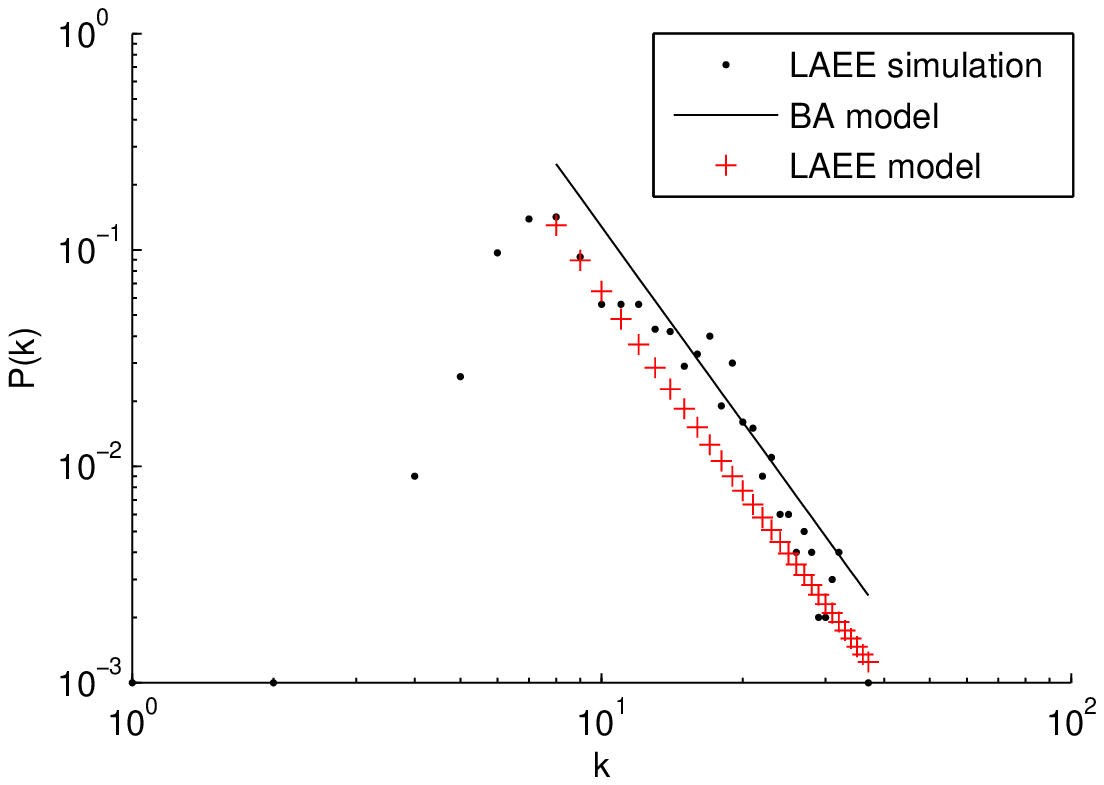}}
\caption{LAEE degree distribution with $ m = $ 3, 5, 8}
\label{Fig2} 
\end{figure}

The simulation and theoretical degree distributions of LAEE  are presented in Fig. \ref{Fig2}. The theoretical degree distribution of LAEE model is close to the degree distribution of BA model, and the simulation result of degree distribution is close to the theoretical value when $ k \geqslant m $. It is noteworthy that the degree values must be large than $ m $ in BA model (each node has $m$ links at least), while there are nodes with degrees less than $ m $ in LAEE simulation results. This is because in our model a node may have the number of potential neighbors less than $m$. If this happens, this node's degree may keep in a low value. In other words, it is due to WSNs' spatial structure. Fortunately, only a small proportion of nodes have degrees less than $ m $. The power-law degree distribution is valid for most of nodes.

\begin{table}[!htb]
\begin{flushleft}
\caption{\label{table1}Parameters for simulation}
\small
\begin{tabular}{ccc}
\toprule
 Parameter		&	Value			&		Definition					\\
\midrule
 $ n $			&	1000			&Number of nodes					\\
 $ r $				&	100$ m $		&Transmission range					\\
 $ S $			&	1000$\times$1000$m^2$	&Entire coverage region		\\
 $ m_0 $			&	10			&Number of nodes in the initial state		\\
 $ e_0 $			&	10			&Number of links in the initial state		\\
 $ m $			&	3, 5, and 8		&Links added in every time step			\\
 $ E_{min} $		&	0.5 $J$		&Lower bound of energy $E$			\\
 $ E_{max} $		&	1 $J$		&Upper bound of energy $E$			\\
 $ K_{max} $		&	30			&Upper bound of degree				\\
 $ \rho $			& $\int_{E_{min}}^{E_{max}}\rho dE=1$  &Uniform distribution of energy $E$	\\
 \bottomrule
\end{tabular}
\end{flushleft}
\end{table}

\begin{table}[!htb]
\begin{flushleft}
\caption{\label{table2}Degrees in topology}
\small
\begin{tabular}{cccc}
\toprule
 Model	 		&	Avg. degree 	& Min. degree 		& Max. degree			\\
\midrule
 UDG			&	29.21		&	6			&	53				\\
 LAEE ($m=3$)		&	5.22			&	1			&	25				\\
 LAEE ($m=5$)		&	7.88			&	1			&	29				\\
 LAEE ($m=8$)		&	11.34			&	2			&	29				\\
 KNN ($k=6$)		&	6			&	6			&	6				\\
 DTG				&	5.85			&	3			&	11				\\	
 LEACH+KNN ($k=6$)&	5.98			&	4			&	6				\\
 LEACH+DTG		&	5.64			&	3			&	10				\\
\bottomrule
\end{tabular}
\end{flushleft}
\end{table}

\begin{figure*}[htbp]
  \includegraphics[width=1\textwidth]{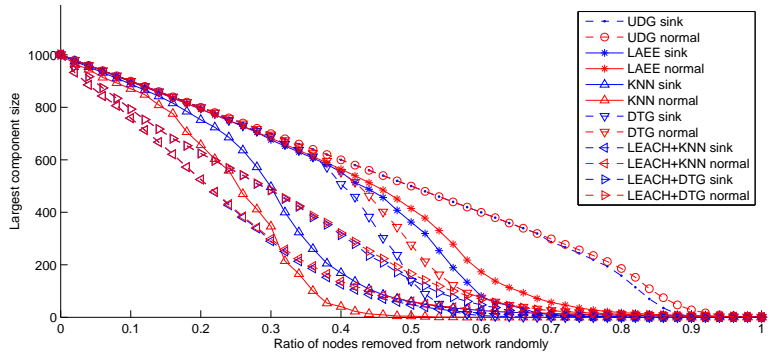}
  \caption{Giant component size of WSNs under random nodes failure}\label{Fig3}
\end{figure*}

Fault tolerance is a key issue in WSNs. Many real applications do not require all nodes to be connected. It is appropriate to consider relaxing the connectivity requirement \cite{XTa2009}. When a fraction of nodes are out of work, the remains may not be connected， and the application of entire network may become invalid. Then, it is important to introduce the giant component, which means the largest connected component \cite{GGanesan2013,HekmatMNA2006}, to measure the fault tolerance of WSNs with the nodes' failure. Two types of data flows exist in WSNs, flows between any pair of nodes, and between sink and other nodes.  Therefore, two kinds of giant component are considered correspondingly: the normal one which contains the largest number of nodes, and the one with the sink. Sometimes these two giant components are the same, whereas sometimes they are not.

We examine how the fault tolerance of WSNs can be improved by LAEE. Nodes are removed randomly to simulate the procedure of energy depletion or random failure. Typical WSNs construction models UDG, KNN,  DTG, LEACH+KNN(LEACH for cluster heads election, KNN for topology construction in each cluster), LEACH+DTG (DTG for topology construction in each cluster) are used for comparison. The degree parameters are shown in Table \ref{table2}. We can see that their average degrees are close to that of LAEE with $ m = 3 $. Close average degrees means these topologies contain similar number of links. However, due to the scale-free feature, the degree distribution of LAEE is much wider than other construction models. As Fig. \ref{Fig3} shows, with the removing of nodes randomly and gradually, the sizes of giant components decrease. UDG provides upper bounds of giant components. The LAEE presents a larger giant components than KNN, DTG, LEACH+KNN, LEACH+DTG, though it has the minimum number of average degree. Therefore we deem that LAEE, which presents the scale-free feature in degree distribution, has better tolerance against energy depletion or random failure in WSNs.

\section{Conclusions}
Topology control is one of the primary challenges to make WSNs resource efficient. In this paper, we propose a local information and energy-efficient based topology evolution model. The process of topology evolution is divided into two phases. In the first phase, nodes are distributed randomly in a fixed region. In the second phase, topology evolution starts from sink, grows with preferential attachment rule, and stops until all nodes are added into network. The theoretical degree distribution of LAEE evolution model is approaching that of BA model. Simulation result shows that when  $ k \geqslant m $, the degree distribution follows the power law. The LAEE model has better tolerance against energy depletion or random failure than other non-scale-free WSNs topology with close average degrees.



\bibliographystyle{spmpsci}      

\bibliography{cssp_jlr}   

%


\end{document}